\documentclass[12pt]{article}
\usepackage{latexsym}
\usepackage{amsmath,,calrsfs}
\usepackage{amsfonts}
\usepackage{amssymb}
\usepackage{amscd}
\usepackage{bbm}
\usepackage{fancybox}
\usepackage{cite}
\usepackage{amsmath,amsfonts,amsbsy}
\usepackage{pstricks,pst-node}
\usepackage[small,bf,hang]{caption2}
\usepackage{graphicx}
\usepackage{epsfig}
\usepackage{psfrag}
\usepackage{comment}

\usepackage{float}

\psset{unit=1.3cm,linewidth=.5pt,radius=.2} % controls the size of diagrams

\usepackage{multirow}           % tables cells spanning multiple rows
\usepackage{float}             % to force floaters to stay Here
\usepackage{lscape}             % landscape laid out pages
\usepackage{bm}

\newcommand{\beq}{\begin{equation}}
\newcommand{\eeq}{\end{equation}}
\newcommand{\bi}{\begin{itemize}}
\newcommand{\ei}{\end{itemize}}
\newcommand{\bt}{\begin{tabular}}
\newcommand{\et}{\end{tabular}}
\newcommand{\bc}{\begin{center}}
\newcommand{\ec}{\end{center}}

\newcommand{\be}{\begin{equation}}
\newcommand{\ee}{\end{equation}}
\newcommand{\bea}{\begin{eqnarray}}
\newcommand{\eea}{\end{eqnarray}}
\newcommand{\ba}{\begin{array}}
\newcommand{\ea}{\end{array}}

\def\bbox{{\,\lower0.9pt\vbox{\hrule \hbox{\vrule height 0.2 cm
\hskip 0.2 cm \vrule height 0.2 cm}\hrule}\,}}
\newcommand{\dsl}{\pa \kern-0.5em /}

\def\tr{{\rm tr}}
\makeatletter \@addtoreset{equation}{section} \makeatother

\def\slashchar#1{\setbox0=\hbox{$#1$}      % set a box for #1
  \dimen0=\wd0                 % and get its size
  \setbox1=\hbox{/} \dimen1=\wd1        % get size of /
  \ifdim\dimen0>\dimen1            % #1 is bigger
   \rlap{\hbox to \dimen0{\hfil/\hfil}}   % so center / in box
   #1                    % and print #1
  \else                    % / is bigger
   \rlap{\hbox to \dimen1{\hfil$#1$\hfil}}  % so center #1
   /                     % and print /
  \fi}

\begin{document}

\begin{titlepage}%1
\begin{center}

\hfill DAMTP-2015-2

\vskip 1.5cm

{\Large \bf On Solutions of Minimal Massive 3D Gravity}\\

\vskip 1cm

{\bf Alex S. Arvanitakis} \\

\vskip 1cm

Department of Applied Mathematics and Theoretical Physics,\\ Centre for Mathematical Sciences, University of Cambridge,\\
Wilberforce Road, Cambridge, CB3 0WA, U.K.\

\vskip 5pt

{email: {\tt A.S.Arvanitakis@damtp.cam.ac.uk}} \\

\end{center}

\vskip 0.5cm

\begin{abstract}
We look at solutions of Minimal Massive Gravity (MMG), a generalisation of Topologically Massive Gravity (TMG) that improves upon its holographic properties. It is shown that generically (in MMG parameter space) all conformally flat solutions of vacuum MMG are locally isometric to one of the two (A)dS vacua of the theory. We then couple a scalar field, and find that domain wall solutions can only interpolate between these two vacua precisely when the bulk graviton is tachyonic. Finally, we find a non-BTZ AdS black hole solution satisfying Brown-Henneaux boundary conditions, which lies within the ``bulk/ boundary unitarity region''.
\end{abstract}

\end{titlepage}

\newpage
\setcounter{page}{1} 
%\tableofcontents

\newpage

%%%%%%%%%%%%%%%%%%%%%%%%%%%
\section{Introduction}
The theory of Minimal Massive Gravity (MMG) in three dimensions put forward in \cite{Bergshoeff2014} is an extension of the older theory of Topologically Massive Gravity (TMG) \cite{Deser1981wh} which maintains the latter's bulk features but has improved boundary behaviour. TMG consistently propagates a single mode around any background; this may be identified as a massive graviton if the background is flat/(A)dS. However, upon fixing AdS boundary conditions, the central charges of the asymptotic symmetry algebra were found to be positive only when the bulk graviton has negative energy, and vice-versa. This has been dubbed the ``bulk/boundary unitarity clash''.

MMG also propagates a single mode around any background, but manages to evade this bulk/boundary unitarity clash when its parameters lie inside the ``unitarity region'' identified in \cite{Bergshoeff2014} and \cite{Arvanitakis:2014xna}. Within that region positive energy bulk gravitons peacefully coexist with positive Virasoro central charges of the putative CFT dual. Thus the possibility arises that unitary MMG might make sense as a holographic theory of massive gravity.

In this work we will investigate the space of MMG solutions. Other works on this topic include \cite{Giribet:2014wla} and \cite{Alishahiha:2014dma}. We will primarily focus on conformally flat spacetimes. It is a trivial observation that any conformally flat solution of TMG is locally flat/(A)dS in the absence of matter. After some work, we will find that this is also true of MMG for generic values of its parameters, a set which includes the entirety of the unitarity region. We also briefly examine the exceptional point in MMG parameter space and discuss some of its properties.

We then couple the theory to scalar matter according to the prescription of \cite{Arvanitakis:2014yja}. We only investigate solutions with a 2D Poincar\'e isometry group for simplicity; a priori this still allows interesting solutions in the form of black strings and domain walls. We find that domain walls that interpolate between the two maximally symmetric MMG vacuum solutions may only exist if the bulk graviton is tachyonic on at least one of the vacua. As for black strings, we will rule out their existence whenever the scalar field potential is trivial.

In the last section of this paper we switch gears and study a non-conformally-flat geometry that describes a black hole in AdS. We determine the conditions under which this geometry appears as a solution of MMG in the absence of matter. Interestingly, it solves MMG {\it within} the unitarity region.

\section{MMG preliminaries}
Here we collect some basic facts about MMG. Our conventions match \cite{Bergshoeff2014} and \cite{Arvanitakis:2014yja}. The MMG equation of motion depends on four parameters $(\mu, \bar{\sigma}, \bar{\Lambda}_0, \gamma)$. If we include coupling to matter it reads
\be
\label{mmgfieldeq}
\frac{1}{\mu} C_{\mu\nu} +\bar{\sigma}G_{\mu\nu} +\bar{\Lambda}_0 g_{\mu\nu} + \frac{\gamma}{\mu^2} J_{\mu\nu} ={\cal T}_{\mu\nu},\qquad J_{\mu\nu}\equiv -\frac{1}{2} \epsilon_\mu^{\ \rho \sigma} \epsilon_\nu^{\ \tau \eta} S_{\rho \tau} S_{\sigma \eta}
\ee
where the tensors $C_{\mu\nu}$ and $S_{\mu\nu}$ are known as the Cotton and Schouten tensors respectively, and are defined by
\bea
S_{\mu\nu} &\equiv& R_{\mu\nu} - \frac{R}{4} g_{\mu\nu} \\
C_{\mu\nu} &\equiv& \epsilon_\mu^{\ \rho \sigma} D_\rho S_{\sigma \nu}\,.
\eea
MMG is a generalisation of the older model of Topologically Massive Gravity (TMG), to which it degenerates at $\gamma=0$. $\cal T_{\mu\nu}$ is a source tensor describing the interaction to matter; it is {\it not} the matter stress-energy tensor $T_{\mu\nu}$, but a quantity derived from it \cite{Arvanitakis:2014yja}. $\cal T_{\mu\nu}$ does, however, reduce to the matter stress-energy tensor at $\gamma =0$.

Setting $\cal T_{\mu\nu}$ to zero for the moment, we notice a peculiarity of the MMG field equation: It does not satisfy a Bianchi identity. One can calculate
\be
D_\mu J^{\mu\nu}= \epsilon^{\nu\rho \sigma} S_\rho^\tau C_{\sigma \tau} \neq 0\,.
\ee
This has numerous implications, the most striking of which is the fact that the theory has no (local, diffeomorphism-invariant) action in terms of the metric alone. This is not to say that the MMG field equations cannot be derived from an action, but rather that that action necessarily contains auxiliary fields which cannot be eliminated in a manner consistent with the variational principle. We refer the reader to \cite{Bergshoeff2014} and \cite{Arvanitakis:2014yja} for detailed discussions on this point.

% A detailed derivation of the MMG equation of motion gives $(\bar{\sigma}, \bar{\Lambda}_0, \gamma)$ as complicated functions of the more fundamental MMG action parameters \cite{Bergshoeff2014}. Since this paper is about MMG solutions we will attempt to avoid using the latter as much as possible. However one of the MMG action parameters, $\alpha$, always appears in expressions involving $\cal T_{\mu\nu}$ so we shall describe it here. $\alpha$ is the MMG action counterpart to $\gamma$: when it vanishes, the MMG action reduces to the TMG action. We have
% \be
% \alpha =0 \iff \gamma =0, \qquad \quad    +\infty>-\frac{\gamma}{\alpha} >0 \ \forall \alpha
% \ee
% and in particular
% \be
% -\left.\frac{\alpha}{\gamma}\right|_{\alpha=0} =1\,.
% \ee
% These expressions can be used to relate our MMG results to the corresponding expressions for TMG.

The simplest vacuum solutions to the MMG field equations (\ref{mmgfieldeq}) are the maximally symmetric spaces, i.e. Minkowski or (A)dS depending on the value of the cosmological constant $\Lambda$. These solutions satisfy
\be
G_{\mu\nu} + \Lambda g_{\mu\nu} =0
\ee
and using this in (\ref{mmgfieldeq}) gives a quadratic equation for $\Lambda$:
\begin{equation}\label{Lambdaeq}
\bar\Lambda_0 - \bar\sigma \Lambda + \frac{\gamma}{4\mu^2} \Lambda^2 =0\, . 
\end{equation}
In the case of TMG $(\gamma =0)$ there is a unique value for $\Lambda$ equal to $\bar\Lambda_0/\bar\sigma$. Otherwise we have {\it two} possible values
\begin{equation}\label{mp}
\Lambda_\pm= \frac{2\mu}{\gamma} \left[\mu\bar\sigma \pm m\right] \, , \qquad m\equiv \sqrt{\mu^2\bar\sigma^2 - \gamma\bar\Lambda_0} \,.
\end{equation}
$m\geq 0$ is required for the existence of maximally symmetric vacua. When this inequality is saturated $\Lambda_+=\Lambda_-$; this is known as the ``merger point'' in MMG parameter space \cite{Arvanitakis:2014yja}.
\section{Conformally flat vacuum spacetimes}
A (locally) conformally flat spacetime has a vanishing Cotton tensor $C_{\mu\nu}$ \cite{Garcia:2003bw}. In the absence of matter fields, the source tensor ${\cal T}_{\mu\nu}$ also vanishes, and the necessary and sufficient condition for such a spacetime to solve MMG is
\be
\label{confflateq}
\bar{\sigma}G_{\mu\nu} +\bar{\Lambda}_0 g_{\mu\nu} + \frac{\gamma}{\mu^2} J_{\mu\nu} =0 \,.
\ee
In this section we will prove that {\it away from the merger point $m=0$ all such solutions are locally maximally symmetric}. This is only non-trivial when $\gamma \neq 0$.

We shall employ a direct approach. Consider the tensor $X_{\mu\nu}$ defined by
\be
X_{\mu\nu}\equiv G_{\mu\nu} + \Lambda_+ g_{\mu\nu}\,.
\ee
Any maximally symmetric spacetime has either
\be
X_{\mu\nu} =0 \qquad \text{or}\qquad X_{\mu\nu}=4\frac{m\mu}{\gamma} g_{\mu\nu}
\ee
corresponding to the two admissible values of the cosmological constant. We will proceed by showing that these are in fact the {\it only} values of $X_{\mu\nu}$ that solve (\ref{confflateq}) away from $m=0$. After writing both $G_{\mu\nu}$ and $S_{\mu\nu}$ in terms of $X_{\mu\nu}$ (\ref{confflateq}) becomes
\begin{align}
0&=\left(\bar{\Lambda}_0- \bar{\sigma} \Lambda_+ +  \frac{\gamma}{4 \mu^2} \Lambda_+^2\right) g_{\mu\nu} \nonumber\\
&+ \left(- \frac{\gamma}{2 \mu^2} \Lambda_+ + \bar{\sigma}\right) X_{\mu\nu} + \frac{\gamma}{\mu^2}X_{\mu\rho}X_\nu^{\ \rho} -  \frac{\gamma}{2\mu^2} X_{\mu\nu} X -  \frac{\gamma}{2\mu^2} g_{\mu\nu} X_{\rho\sigma} X^{\rho\sigma} + \frac{\gamma}{4\mu^2} \
g_{\mu\nu} X^2\,.
\end{align}
The first term is proportional to the quadratic equation for the cosmological constant $\Lambda_\pm$ and thus vanishes, while the second is just $-m/\mu X_{\mu\nu}$.

At this point it is convenient to raise one index and switch to matrix notation. Writing $\pmb X$ for $X_{\mu\rho}g^{\rho\nu}$ and $\pmb I$ for $\delta^\mu_\nu$ gives
\be
\label{matrixeq}
0=-\frac{m}{\mu} {\pmb X} + \frac{\gamma}{\mu^2} \pmb{X}^2 - \frac{\gamma}{2\mu^2}\tr(\pmb{ X}) {\pmb X}- \frac{\gamma}{2\mu^2}  \tr\left(\pmb{X}^2\right)\pmb I+\frac{\gamma}{4\mu^2}\tr(\pmb{ X})^2 \pmb I\,.
\ee
The eigenvalue equation for $\pmb X$ is easily solved once one notices that it is quadratic, and hence there are at most two distinct eigenvalues rather than three. The solutions for the eigenvalues $\lambda_i,\, i=1,2,3$ are
\begin{itemize}
\item For $m\neq 0:\, \lambda_1=\lambda_2=\lambda_3\equiv \lambda,\qquad \lambda=0\qquad \text{or}\qquad \lambda = 4m\mu/\gamma \ $
\item For $m=0:\qquad
\lambda_1=0\qquad \text{and}\qquad \lambda_2=\lambda_3\qquad$(and permutations thereof)
\end{itemize}
Leaving aside the case $m=0$, and assuming $\lambda =0$ for definiteness, we can without loss of generality write $\pmb X$ in the form
\be
\pmb X = \begin{pmatrix} 0 & \psi_1 & \psi_2 \\ 0 &0 &\psi_3\\ 0&0&0 
\end{pmatrix}\,.
\ee
Equation (\ref{matrixeq}) for $\pmb X$ then reads very simply
\be
\label{matrixeq2}
0=\begin{pmatrix} 0 & -\frac{m}{\mu}\psi_1 & -\frac{m}{\mu}\psi_2 + \frac{\gamma}{\mu^2} \psi_1 \psi_3 \\ 0 &0 &-\frac{m}{\mu}\psi_3\\ 0&0&0 \,.
\end{pmatrix}
\ee
This has the unique solution $\psi_i=0$ and thus {\it all conformally flat solutions of MMG are locally maximally symmetric away from $m=0$.} The case $\lambda=4m\mu/\gamma$ works similarly.

An easy corollary is that all solutions with a hypersurface orthogonal Killing vector---a class which includes static spacetimes---are also locally maximally symmetric. This follows from the fact that there is a complete mismatch between the parallel and orthogonal components of the Ricci and Cotton tensors \cite{Aliev:1996eh,Deser:2009er}, implying that the Cotton tensor has to vanish.
\subsection{The ``merger point''}
The ``merger point'' in MMG parameter space was first identified in \cite{Arvanitakis:2014yja} as the point in $(\gamma, \bar{\sigma}, \bar{\Lambda}_0)$ space where the two possible values of the cosmological constant of a maximally symmetric vacuum coincide. It can also be characterised in other ways:
\begin{itemize}
\item As the point where the theory admits ``Kaluza-Klein'' vacua of the form (A)dS$_2\times$S$^1$ and the attendant non-BTZ black hole solutions \cite{Arvanitakis:2014yja}, and
\item as the point where the mass parameter $M$ of linearised MMG vanishes \cite{Bergshoeff2014}, and the theory has a linearisation instability \cite{Arvanitakis:2014xna}\cite{Alishahiha:2014dma}.
\end{itemize}
These observations are somewhat puzzling, in that they involve disparate phenomena of both the linearised and the full non-linear theory. Here we shed light on this issue by identifying the merger point in yet another manner, as the point in parameter space where the Einstein and cosmological terms of the MMG field equation (\ref{mmgfieldeq}) can be ``absorbed'' into the J-tensor by a shift of $S_{\mu\nu}$.

What we mean by ``absorbing'' the Einstein and cosmological terms into $J_{\mu \nu}$ is the following. One first makes the replacement
\be
S_{\mu \nu} \to S'_{\mu\nu} \equiv S_{\mu\nu} + C_1 g_{\mu\nu}
\ee
in the definition of J, obtaining
\begin{align}
J'_{\mu\nu} &\equiv -\frac{1}{2} \epsilon_\mu^{\ \rho \sigma} \epsilon_\nu^{\ \tau \eta} S'_{\rho \tau} S'_{\sigma \eta} \nonumber \\
&=J_{\mu\nu} + C_1 \left[ g_{\mu\nu} S - S_{\mu\nu} \right] + C_1^2 g_{\mu\nu} \nonumber \\
&=J_{\mu\nu} - C_1G_{\mu\nu} + C_1^2 g_{\mu\nu} \label{jprime} \,.
\end{align}
It is then a simple matter to solve $ \frac{\gamma}{\mu^2} J'_{\mu\nu} = \bar{\sigma}G_{\mu\nu} +\bar{\Lambda}_0 g_{\mu\nu} + \frac{\gamma}{\mu^2} J_{\mu\nu}$ for $C_1$. The result is
\be
C_1 = - \frac{\mu^2\bar{\sigma}}{\gamma}\qquad \text{with} \qquad \gamma \bar{\Lambda}_0 - (\mu \bar{\sigma})^2 =0
\ee
where the second equality is the merger point condition. $C_1$ now has a simple interpretation; it is just $-\Lambda/2$. To sum up, at (and only at) the merger point we can rewrite the MMG field equation in the very compact form
\be
0=\frac{1}{\mu} C_{\mu\nu}-\frac{\gamma}{2 \mu^2} \epsilon_\mu^{\ \rho \sigma} \epsilon_\nu^{\ \tau \eta} S'_{\rho \tau} S'_{\sigma \eta}, \qquad S'_{\mu\nu} \equiv S_{\mu\nu} -\frac{\Lambda}{2} g_{\mu\nu} \,.
\ee

A maximally symmetric background of cosmological constant $\Lambda$ satisfies $G_{\mu\nu} = - \Lambda g_{\mu\nu}$ or $S'_{\mu\nu} =0$. It is now easy to see the origin of the linearisation instability: $J'_{\mu\nu}$ vanishes to {\it quadratic} order around any maximally symmetric background, and the linearised equation of motion is just
\be
\tilde{C}_{\mu\nu} =0
\ee
where $\tilde{C}_{\mu\nu}$ is the linearised Cotton tensor. This is the linearised equation of motion for the theory of 3D Conformal Gravity \cite{Horne:1988jf}, which is invariant under local rescalings of the metric. Therefore we have an extra gauge invariance of linearised MMG at the merger point---the linearised Weyl rescaling of the metric---on top of the usual linearised diffeomorphisms, reducing the number of propagating degrees of freedom from one to zero as far as the linearised theory is concerned.

This ``factorised'' form of the equation of motion is also responsible for the wider variety of MMG solutions at the merger point. The salient point is that there exist nontrivial ans\"atze for $S'_{\mu\nu}$ which make both $J'_{\mu\nu}$ and $C_{\mu\nu}$ vanish identically.

As an example, take
\be
\label{xansatz}
S'_{\mu\nu} =A V_\mu V_\nu
\ee
which makes $J'_{\mu\nu}$ vanish automatically by antisymmetry. It also satisfies $C_{\mu\nu}=0$ if for instance A is constant and $V_\mu$ is covariantly constant, or
\be
D_\mu V_\nu =0\,.
\ee
This property implies that $V$ is Killing and hypersurface-orthogonal. 

In a coordinate system adapted to $V$, the metric and Ricci tensor completely factorise into a 1-dimensional parallel and a 2-dimensional perpendicular block, and the equation (\ref{xansatz}) can easily be solved to give solutions of ``Kaluza-Klein'' form, i.e. product spacetimes with two maximally symmetric factors. If for instance $V$ is taken to be spacelike and to have closed integral curves then the solutions are exactly the (A)dS$_2\times$S$^1$ vacua found in \cite{Arvanitakis:2014yja}.

\section{Domain walls and black strings}
Let us now consider MMG geometries supported by a scalar field. For simplicity, we focus on spacetimes with a 2D Poincar\'e isometry group. This is a particular subclass of conformally flat spacetimes in three dimensions. The metric can always be written
\be
\label{2dpoincareisomansatz}
ds^2=a(r)^2 (-dt^2 +dx^2) + dr^2\,,
\ee
at least locally, and can describe the geometry of either a domain wall or a black string. We shall investigate both possibilities.

% Metrics of this form can describe two physically distinct scenarios, depending on the asymptotics. If a metric of the form (\ref{2dpoincareisomansatz}) is asymptotic to a maximally symmetric (flat or AdS) vacuum of cosmological constant $\Lambda$ at both $r\to + \infty$ and $r\to - \infty$ it is interpreted as an {\it extremal black string} in that vacuum spacetime. If a metric of the form (\ref{2dpoincareisomansatz}) is asymptotic to two {\it different} vacua, say $\Lambda_1$ at $r \to +\infty$ and $\Lambda_2$ at $r \to -\infty$ the physical interpretation is different: then, the metric (\ref{2dpoincareisomansatz}) describes a {\it domain wall} that interpolates between the two vacua.

To this geometry we couple a self-interacting scalar field $\Phi$ with action
\be
I_{\rm matter}= \int d^3x  \ \sqrt{-g} \left\{ -\frac{1}{2}g^{\mu\nu} D_\mu \Phi D_\nu \Phi -V(\Phi) \right\}
\ee
and Belinfante-Rosenfeld stress-energy tensor
\be
T_{\mu\nu}=D_\mu \Phi D_\nu \Phi -\frac{1}{2}g_{\mu\nu}\left( D_\rho\Phi D^\rho \Phi + 2 V(\Phi) \right)\,.
\ee
According to the prescription of \cite{Arvanitakis:2014yja}, the source tensor $\cal T_{\mu\nu}$ that enters the MMG field equations (\ref{mmgfieldeq}) depends on the matter fields only through this $T_{\mu\nu}$. We refer to that paper for the explicit formula.

A detailed derivation of the MMG equation of motion gives $(\bar{\sigma}, \bar{\Lambda}_0, \gamma)$ as complicated functions of the more fundamental MMG action parameters \cite{Bergshoeff2014}. Since this paper is about MMG solutions we have avoided using the latter as much as possible. However one of the MMG action parameters, $\alpha$, always appears in expressions involving $\cal T_{\mu\nu}$ so we shall describe it here. $\alpha$ is the MMG action counterpart to $\gamma$: when it vanishes, the MMG action reduces to the TMG action. We have
\be
\alpha =0 \iff \gamma =0, \qquad \quad    +\infty>-\frac{\gamma}{\alpha} >0 \ \forall \alpha
\ee
and in particular
\be
-\left.\frac{\alpha}{\gamma}\right|_{\alpha=0} =1\,.
\ee
These expressions can be used to relate our MMG results to the corresponding expressions for TMG.

The equation of motion of the scalar in the coordinate system of (\ref{2dpoincareisomansatz}) can be written
\be
2 a' \Phi' + a( \Phi'' - dV/d\Phi)=0
\ee
and the $rr$ component of the MMG field equation (\ref{mmgfieldeq}), after considerable rearranging, is
\be
\frac{1}{4 \mu^2} \left[ \left(\frac{a'}{a}\right)^2 + \frac{\alpha}{2} \left(  (\Phi')^2-2V(\Phi)\right)+ \frac{2 \mu^2 \bar{\sigma}}{\gamma} \right]^2 = \frac{1}{\gamma^2} (\mu^2 \bar{\sigma}^2-\gamma \bar{\Lambda}_0)-\frac{\alpha}{2 \gamma^2} \left[(\Phi')^2-2V(\Phi) \right]\,.
\ee
It can be verified that when both of these hold, the full equations of motion are satisfied.

These equations simplify considerably after defining $\theta= (\Phi')^2-2 V$ and $\lambda \equiv 2 \mu^2 \bar\sigma/\gamma$. If we also take a square root in the second equation we obtain
\begin{align}
\label{athetarel}\frac{a'}{a} &= -\frac{1}{4} \frac{\theta'}{(\theta + 2 V)}\\
\label{rreq}\left(\frac{a'}{a}\right)^2&= - \frac{\alpha \theta}{2} - \lambda \pm \frac{2 \mu}{\gamma} \sqrt{m^2 - \frac{\alpha \theta}{2} }
\end{align}
and using (\ref{athetarel}) in (\ref{rreq}) gives a single equation for $\theta$:
\be
\label{thetaeq}
\frac{1}{16}\left(\frac{\theta'}{(\theta + 2 V)}\right)^2= - \frac{\alpha \theta}{2} - \lambda \pm \frac{2 \mu}{\gamma} \sqrt{m^2 - \frac{\alpha \theta}{2} }
\ee
In the absence of matter the sign in the second equation determines the value of the cosmological constant to be $\Lambda_\mp$ as defined in (\ref{mp}).

From this point on we will treat domain walls and black strings separately.

\subsection{Domain walls}
By {\it domain wall} we denote a metric of the form (\ref{2dpoincareisomansatz}) which is asymptotic to two MMG vacua. We will first investigate the possibility of interpolating between the two maximally symmetric vacua (\ref{mp}) with cosmological constants $\Lambda_+$ and $\Lambda_-$ (both necessarily nonpositive). We will make the assumptions
\be
\lim_{|r|\to \infty} \Phi(r)=0, \qquad V(0)=0, \qquad \left.\frac{d V}{d\Phi}\right|_{\Phi=0} =0\,.
\ee
The first two assumptions ensure that the asymptotic geometry is an MMG vacuum of cosmological constant $\Lambda_\pm$, while the last one follows from consistency of the scalar field equation in the asymptotic region. We will also assume that the potential is appropriately differentiable.

Given these assumptions, the equation (\ref{thetaeq}) must be satisfied with {\it both} signs, one for each asymptotic region. Therefore if everything is continuous, there exists a value of $r$ where the quantity in the square root vanishes, i.e.
\be
 \theta= \frac{2m^2}{\alpha}\,.
\ee
This is only consistent with (\ref{rreq}) when
\be
m^2 + \lambda\leq 0\,.
\ee
From (\ref{mp}) we observe that $\lambda$ is the average of $\Lambda_+$ and $\Lambda_-$. If it is negative, then one of
\be
{\cal M}^2_\pm\equiv \Lambda_\pm + m^2
\ee
is also negative, where ${\cal M}_\pm$ is the mass of the MMG bulk graviton mode on the (A)dS background with cosmological constant $\Lambda_\pm$. \cite{Bergshoeff2014}. Therefore, domain walls relating the two MMG vacua only exist whenever the bulk gravitons are tachyonic. In particular, they do not exist in the ``unitarity region'' found in \cite{Bergshoeff2014} and \cite{Arvanitakis:2014yja}.

At the merger point $m=0$, $\Lambda_+ = \Lambda_-$ and so domain walls of the above type cannot exist. Therefore we will look for solutions which are asymptotic to the same AdS vacuum of cosmological constant $\Lambda=\lambda$ in both directions, and thus lift the restriction that (\ref{thetaeq}) hold with different signs in each asymptotic region. Interestingly, (\ref{thetaeq}) with $m=0$ is only consistent if $\alpha <0$ (equivalently, $\gamma >0$); this is in contrast to the situation with cosmological spacetimes \cite{Arvanitakis:2014yja}.

If $V=0$ (\ref{thetaeq}) reads simply
\bea
\left(\frac{u'}{u}\right)^2 = 4 (u-u_1)(u-u_-),\qquad u\equiv \sqrt{-\frac{\alpha}{2} \theta},\\
u_1 \equiv \mp \frac{\mu}{\gamma} + \frac{\sqrt{\gamma^2 \Lambda + \mu^2}}{\gamma}, \qquad u_2 \equiv \mp \frac{\mu}{\gamma} - \frac{\sqrt{\gamma^2 \Lambda + \mu^2}}{\gamma}
\eea
where $\mp$ is correlated with the sign in $(\ref{thetaeq})$. The solution only exists and is nonsingular when $(\gamma^2 \Lambda + \mu^2 >0, \Lambda<0)$ and can be written down in closed form:
\be
u(r) = \frac{-\gamma \Lambda}{\sqrt{\gamma^2 \Lambda + \mu^2} \cosh\left( 2\sqrt{-\Lambda} r\right) + \mu}\,.
\ee
The signs have been selected to avoid unacceptable singularities and we also set $r=0$ at $u=u_{\rm max}$ for convenience. Using (\ref{athetarel}) gives
\be
a(r)^2 \propto \frac{\sqrt{\gamma^2 \Lambda + \mu^2} \cosh\left( 2\sqrt{|\Lambda|} r\right) + \mu}{-\gamma \Lambda}\,.
\ee
This is identical to the explicit cosmological ``Big Bounce'' solution of \cite{Arvanitakis:2014yja} with $\Lambda \to -\Lambda$, $t \to r$ and $\kappa=1$.

\subsection{Black strings}
In order for a metric of the form (\ref{2dpoincareisomansatz}) to describe a (necessarily extremal) {\it black string}, there must exist a Killing horizon, i.e. a point where $a(r)$ vanishes. At the same time $(a'/a)^2$ is finite on the horizon; it is equal to minus the value of the Ricci scalar of the near-horizon geometry, which for a black string in 3D is AdS$_3$.

If we assume $V=0$ so that the equations can be integrated explicitly, (\ref{athetarel}) gives $\theta$ as a function of $a$:
\be
\theta(r)=\frac{a_0^4}{a(r)^4}\,.
\ee
This diverges as $a\to 0$ and thus the right-hand side of (\ref{rreq}) also diverges.

This does not mean that black strings cannot form under less restrictive assumptions. Assuming a positive potential, $\theta$ is only bounded above by $a^{-4}$. Therefore it is not implausible that such solutions exist, although we will not investigate this possibility further here.

\section{A non-BTZ black hole}
We will now demonstrate that when one lifts the assumption of conformal flatness, interesting solutions do appear. The following metric was first found in \cite{Giribet:2011vv} (equations (21) and (23)) as a solution of ``chiral'' Generalised Massive Gravity\footnote{For the corresponding ``Generalised MMG'', see \cite{Setare:2014zea}}. We shall see that it also solves vacuum MMG under less stringent conditions.

If we write the extremal BTZ black hole in the form
\be
ds^2|_{\rm BTZ} = - \left(\frac{r^2}{\ell^2} - 4 \beta + \frac{4 \beta^2 \ell^2}{r^2} \right) dt^2 + \frac{dr^2}{\left(\frac{r^2}{\ell^2} - 4 \beta + \frac{4 \beta^2 \ell^2}{r^2} \right)} +r^2 \left(d \theta +\frac{2 \beta \ell}{r^2} dt \right)^2
\ee
and add to it a finite perturbation
\be
ds^2|_{\rm pert}=\kappa \left( \frac{r^2}{\ell^2} -2 \beta \right)^q (dt + \ell d\theta )^2 \,,
\ee
the combined expression 
\be
\label{bhsol}
ds^2 = ds^2|_{\rm BTZ}+ds^2|_{\rm pert}
\ee
solves MMG under certain conditions on its parameters $(q , \kappa, \beta, \ell)$ in the regions $r>0$ and $r<0$. We refer to \cite{Giribet:2011vv} for properties of the solution; here we will only note that this geometry is stationary and, if $q$ is negative, satisfies Brown-Henneaux boundary conditions.

As one might suspect, a necessary condition for the metric (\ref{bhsol}) to solve the MMG field equations (\ref{mmgfieldeq}) in the absence of matter is that $\ell^2$ has to solve a quadratic equation involving the MMG parameters $(\bar{\sigma},\gamma,\bar{\Lambda}_0)$ that identifies it as the radius of the AdS vacuum this solution asymptotes at infinity. To spell the other conditions out, we first need to specify our conventions: We will assume without loss of generality that $\ell,\mu>0$, and define the orientation by $\epsilon^{t\, r\, \theta} >0$. With these conventions, the metric (\ref{bhsol}) then solves MMG when either of the following conditions hold: 
\begin{itemize}
\item \be q (q-1) \kappa =0\,. \ee Then the above metric actually solves the Einstein equation and thus is locally AdS everywhere.
\item \be \frac{(2q-1)}{\ell} = \text{sgn}(r) M,\qquad M \equiv \frac{\gamma + 2 (\mu \ell)^2 \bar{\sigma}}{2 \mu \ell^2}\ee where $M$ is one of the square roots of $m^2$, given in terms of the MMG action parameters (see \cite{Bergshoeff2014}). We stress that when this condition is satisfied the metric is not Einstein (unless the conditions degenerate).

The sign $\text{sgn}(r)$ arises from the contribution of the Cotton tensor. It may be understood as follows: The regions $r>0$ and $r<0$ are related by an {\it orientation-reversing} diffeomorphism, namely $( t \to t,r\to -r, \theta \to \theta)$, under which the Cotton tensor gains an extra minus sign. This effect can be compensated for by changing the sign of $\mu$, but as we have fixed that to be positive we must consider both signs in $r$ instead.
\end{itemize}

In either case, the solution has constant scalar curvature
\be
R=-\frac{6}{\ell^2}
\ee
and in the second case the Cotton tensor components in the $(t,r,\theta)$ coordinates are all proportional to
\be
\kappa  M\ell 
   \left[(M\ell )^2-1\right]
    \left(\frac{r^2}{\ell ^2}-2
   \beta \right)^{\frac{\text{sgn}(r)M\ell
   +1}{2}}
\ee
where we have solved the equation for $q$ in terms of the MMG mass parameter $M$. This expression vanishes in the asymptotic region precisely when $\text{sgn}(r)M\ell <-1 \iff q< 0$, as expected.

Let us analyse the second case further. The condition $q<0$, necessary for this geometry to asymptote AdS as $|r| \to \infty$, implies
\bea
0>\text{sgn}(r) M \ell,\qquad 0< M^2 + \Lambda \,.
\eea
Both inequalities have physical interpretations. As noted in the previous section, the second inequality is the statement that the bulk graviton has positive mass-squared. According to the analysis of \cite{Bergshoeff2014}, the graviton has positive energy only if $M/\mu<0$. Hence, in our conventions, bulk unitarity selects the metric (\ref{bhsol}) with $r>0$. Since there are no more conditions on the MMG parameters, we can also arrange for the boundary central charges to be positive (this is most easily seen in the parameterisation of \cite{Arvanitakis:2014xna}). Therefore, the metric (\ref{bhsol}) with $r>0$ solves MMG with the MMG parameters lying inside the bulk/boundary unitarity region.

%%%%%%%%%%%%%%%%%%%%%%%%%%%%%%%%%
\section{Discussion}
In this paper we studied the space of solutions to the MMG field equations (\ref{mmgfieldeq}), both by proving general results and by considering specific examples. We proved that away from the merger point $m=0$ all conformally flat MMG solutions actually solve the Einstein equations in 3D, and therefore are locally maximally symmetric. This fact has implications for the quantum theory. The Euclidean partition function is dominated by the saddle points of the Euclidean path integral, which are given by solutions of the Wick rotated equations of motion. As the Cotton tensor is third order in derivatives, it enters the Euclidean equation of motion with an extra factor of $i$ \cite{Maloney:2009ck}. Combined with the fact that the other terms in the MMG field equation are of even orders in derivatives and thus do not acquire any factors of $i$, this implies that any {\it real-valued} saddle point is conformally flat. Thus our result neatly characterises the class of real geometries one needs to sum over.

The other major result of this paper is the existence of the black hole solution (\ref{bhsol}) inside the region in MMG parameter space where the bulk theory is unitary and the central charge of the conjectured CFT dual is positive, called the ``unitarity region''. As an example of a non-BTZ black hole within the unitarity region, this geometry is clearly deserving of further study. It would be of interest to see whether it possesses sensible conserved charges (using e.g. the formalism of \cite{Tekin:2014jna}) and thermodynamics (see \cite{Setare:2015pva}).

%%%%%%%%%%%%%%%%%%%%%%%%%%%%%%%%%%%%
\section*{Acknowledgements}
I would like to thank Chris Halcrow, Rahul Jha, and Alasdair Routh for helpful discussions, and my Ph.D. supervisor Paul Townsend for both suggesting this project and providing valuable feedback.
I acknowledge support from the UK Science and Technology Facilities Council (grant number ST/L000385/1), Clare Hall College, Cambridge, and the Cambridge Trust.

\bibliographystyle{hplain}  
%Call bibitex file
\bibliography{references}
\addcontentsline{toc}{chapter}{Bibliography}

\end{document}